\documentclass[letterpaper, 10 pt, conference]{ieeeconf}  

\IEEEoverridecommandlockouts                              

\usepackage{xcolor}
\usepackage{url}
\usepackage{comment}
\usepackage{graphicx}
\usepackage{caption}
\usepackage{subcaption}
\usepackage{multirow}
\usepackage{svg}
\usepackage{adjustbox}
\usepackage{xcolor,colortbl}
    \definecolor{Gray}{gray}{0.85}

\title{\LARGE \bf
The Effect of Data Visualisation Quality and Task Density on Human-Swarm Interaction}

\author{Ayodeji O. Abioye$^{1}$, Mohammad Naiseh$^{2}$, William Hunt$^{1}$, Jediah Clark$^{1}$, \\ Sarvapali D. Ramchurn$^{1}$, and Mohammad D. Soorati$^{1}$
\thanks{$^{*}$Emails: \{a.o.abioye, w.hunt, j.r.clark, sdr1, m.soorati\}@soton.ac.uk, mnaiseh1@bournemouth.ac.uk}%
\thanks{$^{1}$Electronics and Computer Science, University of Southampton, UK.}%
\thanks{$^{2}$Computing and Informatics, University of Bournemouth, UK.}%
\thanks{$^{**}$The authors wish to acknowledge the support received from the UKRI Trustworthy Autonomous Systems Hub (EP/V00784X/1) and the EPSRC project on Smart Solutions Towards Cellular-Connected Unmanned Aerial Vehicles System (EP/W004364/1).}%
}

\begin{document}

\maketitle
\thispagestyle{empty}
\pagestyle{empty}

\begin{abstract}

Despite the advantages of having robot swarms, human supervision is required for real-world applications. The performance of the human-swarm system depends on several factors including the data availability for the human operators. In this paper, we study the human factors aspect of the human-swarm interaction and investigate how having access to high-quality data can affect the performance of the human-swarm system--- the number of tasks completed and the human trust level in operation. We designed an experiment where a human operator is tasked to operate a swarm to identify casualties in an area within a given time period. One group of operators had the option to request high-quality pictures while the other group had to base their decision on the available low-quality images. We performed a user study with 120 participants and recorded their success rate (directly logged via the simulation platform) as well as their workload and trust level (measured through a questionnaire after completing a human-swarm scenario). The findings from our study indicated that the group granted access to high-quality data exhibited an increased workload and placed greater trust in the swarm, thus confirming our initial hypothesis. However, we also found that the number of accurately identified casualties did not significantly vary between the two groups, suggesting that data quality had no impact on the successful completion of tasks.

\end{abstract}


\section{Introduction}


Large robot swarms are an example of complex systems and efficient interaction with them can be extremely challenging. Swarms typically operate in large numbers, ranging from tens to thousands of individual robots. Coordinating and communicating with such a large group can be overwhelming for humans. Interacting with robot swarms requires an interface that is understandable and transparent in order for humans to trust the system. Transparency can lead to improved trust in multi-robot teams \cite{Fox2017}, and increases situational awareness by providing information about the past (i.e. why the swarm exhibited a behaviour), present (i.e. what its current state is), and future (i.e. what will it do in the near and far future) \cite{Chen2014a}. Transparency in swarm robotics can be expensive and may take on two distinct levels: a micro-level and a macro-level \cite{Hepworth2021}. Micro-level transparency communicates each individual drone’s state and intentions. Whereas macro-level transparency aggregates the robot state information to provide an overall decision aid for the human operator. As swarm sizes increase, micro-level transparency may exceed the operator’s ability to track and manage the states of individual drones \cite{Nunnally2012,Roundtree2019}. Further, as display requirements increase, the risk of increased cognitive overload of the operator can have an impact on efficiency, visual search performance, and ultimately task success \cite{Woods1999,Wu2016a}. In these scenarios, aggregated information may be more appropriate for managing tasks, calibrating trust, and relaying information between human operators and robotic swarms. However, aggregated information, such as an image collected by multiple robots to inform human operators about a casualty within a fire area, may require higher bandwidth to transfer the data to the operators. In high-stakes scenarios, especially in adverse weather conditions where communication may be limited, transferring high-quality data to human operators can be expensive and can cause latency in the human-swarm decision-making process. To address this issue, one solution is to transfer low-quality data to human operators and rely on their expertise to make informed decisions. Studies have demonstrated that humans are adept at analyzing visualized data quickly and efficiently, and can also comprehend the context of data even when it is subject to significant noise. Humans can use heuristics to mitigate data noise and transformation, providing a generalized understanding of the data~\cite{KIRCHNER20061762,CHENG2019104649}. Low-quality data may allow a user to get a holistic picture of the swarm status and scenario with minimal delay in relaying the information and can be useful where communication is limited. For instance, a robot swarm that is exploring an area for search and rescue may send lightweight low-quality images to the human operator. 

Task density consideration is also important in human-swarm interaction to avoid cognitive overload which results in diminishing human performance. According to Hussein and Abbass \cite{Hussein2018}, workload needs to be maintained within an acceptable range since both very low and very high levels of workload can cause human performance degradation.

This paper presents the results of an experiment aimed at examining the impact of data quality on human-swarm interaction during high-stakes decision-making tasks. Specifically, we investigate whether providing high-quality data to human operators can significantly enhance the success of the human-swarm scenario, and assess its effect on the operators' trust, workload, and perception of the swarm's performance during the scenario. In order to establish the connections between these components, quantitative data was collected, incorporating both subjective and objective measurements.


\section{Related Work}
Robotic swarms consist of a distributed cognitive network promising to enhance tasks in terrestrial, aerial, and aquatic environments \cite{Schranz2020,Paas2022swarmimpact}. Robotic swarms are currently being used in a variety of application domains such as agriculture~\cite{Abioye2020}, search and rescue~\cite{Dah-Achinanon2023}, monitoring and patrol~\cite{Abioye2018}, warehouse operation~\cite{Liu2017}, last-mile delivery~\cite{Kaufmann2021}, military scenarios~\cite{Bluhm2021}, environmental monitoring~\cite{Tosik2016}, and space exploration~\cite{Nguyen2019}. The move from multi-UAVs to those of greater numbers is challenging, as the ability of a human operator to manage each drone individually becomes less feasible as the swarm size increases \cite{Roundtree2018,Roundtree2019}.

Research investigating the balance between task completion speed and performance accuracy in order to optimise human-swarm performance is not new \cite{Valentini2014,Hussein2019}. However, works focusing on the interplay between human-robot performance (speed and accuracy) against trust are scarce. Hussein et al \cite{Hussein2019} develop a model to study the combined effect of speed and reliance (accuracy) on trust. Their model was based on an assumption that both speed and reliance were crucial to mission performance. They validated their model by conducting human experiments with 33 subjects and suggested that their model generated data which closely mirrored the human data and can be used in trust experiments. This computational trust model was used in identifying components required for the development of trust-aware human automation interaction \cite{Hussein2020}. Further research showed that reliability and transparency have distinct effects on trust calibration in human-swarm interaction \cite{Hussein2020b}. Trust in human-swarm interaction is thought to decrease as performance degrades \cite{Capiola2022}. The level of human reliance on the system in human-swarm interaction is dependent on the level of trust \cite{Hussein2020a}. In our research, we used trust metrics established in human factors and psychology to determine the effect of the trade-off between speed and accuracy on trust. $120$ participants used the \textit{HARIS} simulator ~\cite{hunt2023demonstrating} to interact with an aerial swarm.

Yin et al. \cite{Yin2019} investigated the effect of stated accuracy and observed accuracy on trust in human-robot interaction. They found that users' perception of accuracy was influenced by the observed accuracy than the stated accuracy. This suggests that humans are likely to trust a system based on their perception of how well it is performing. In our study, we investigate how this performance perception in human-swarm interaction can be affected by the quality of data transmitted. Selah et al. \cite{Salah2020} and Naiseh et al. \cite{Naiseh2021calibratedtrust} warn that highly reliable (accurate) machines tend to cause humans to over-trust the system which is usually exhibited as the complacency that limits the ability to detect and fix machine errors. The researchers argued that designing trust-aware human swarm interaction systems may reduce this negative impact of human reliance. In our research, we establish an empirical and qualitative relationship between performance and trust in the human-swarm interaction system. This could inform future designs of such trust-aware human swarm interaction systems. 

Szafr et al \cite{Szafr2021} investigated the effect of data visualisation in human-robot interaction. They highlighted the fact that data visualisation is essential for effective human decision-making. This requires that robot data be represented in a way that supports fast and accurate analyses by humans. They suggested that a well-designed interface would increase performance speed, system trust, and understanding in detecting anomalous behaviours by the human operator. A survey of literature on human-swarm interaction suggested that information transparency was a key component for setting up effective human-swarm teams~\cite{Dahiya2023}. In a multi-operator and multi-robot swarm study~\cite{Patel2020}, it was found that responsibility overlap and balance task density (workload) increased mutual trust and system performance. They proposed the need for the relationship between humans and robots to be more legible and transparent. Another study investigated ergonomic display of human-swarm interaction and highlighted that a complex situation with high dynamics increases the demand on a user~\cite{Bluhm2021}. 

In our previous work~\cite{Soorati2021}, we showed how different visualisation techniques affect human-swarm interaction. We found that heatmap displays (macro-transparency) were preferred in situations with larger swarm sizes, time-critical applications, and tracking motion or progress coverage, whereas individual drone displays (micro-transparency) were preferred for troubleshooting or detecting errors within the swarm. In this research, we extend our investigation into how detailed a visualisation must be to increase trust and overall system performance.

\section{Method}\label{Section:Method}

Our study aimed to investigate the impact of data quality in the transmitted data between humans and swarms on the subjective perception of the swarm. We focused on three key subjective perceptions: trust, workload, and performance. Additionally, we explored the effects of increased workload (task density) on our independent variables. Finally, we used performance metrics for the human-swarm task to measure the effect of data quality in an objective way. To achieve this, we designed a scenario that involved collaborative work between a human operator and a swarm of robots for searching and identifying casualties in a specific area. We selected search and rescue as the use case scenario based on our previous research~\cite{Clark2022}, as it encompasses data quality, workload, and trust as crucial factors for an effective human-swarm partnership. Our study involved participants completing the same scenario while we controlled the variables of data quality and task density. One group performed the scenario with only low-quality images of casualties, while the remaining participants also had the option to request high-quality images. However, there was a simulated delay in receiving high-quality images to mimic real-world conditions. Participants were required to complete two scenarios, one with a low workload and the other with a high workload. By comparing these conditions, our study aimed to explore the potential effects of data quality and task density on human operators' subjective perception of the swarm and the overall performance of the human-swarm team. We formulated the following hypotheses to guide our investigation: 

\begin{itemize}
    \item H1: The presence of high-quality data would increase participants' trust perception and overall human-swarm performance;
    \item H2: Increasing the task density would decrease trust perception and overall human-swarm performance.
\end{itemize}

These hypotheses suggest that other components could also affect human trust perception of the swarm other than transparency \cite{Hussein2020b}, accuracy \cite{Yin2019}, and interface design \cite{Szafr2021}.

\begin{figure}[!htb]
  \centering
  \framebox{\includegraphics[width=0.45\textwidth]{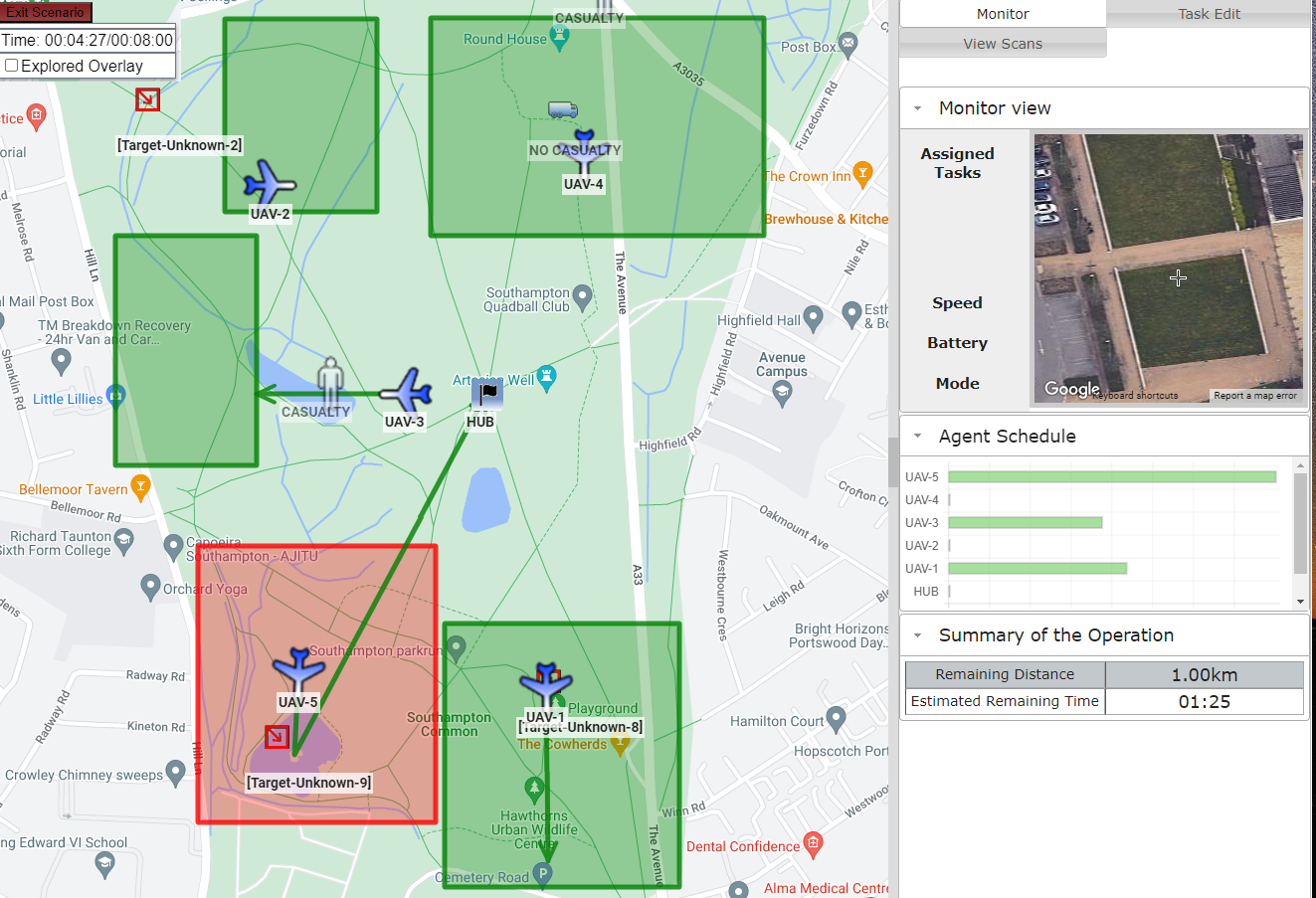}}
  \caption{The operator view allows a human to task the swarm with searching the area for potential casualties}
  \label{fig:OperatorView}
\end{figure}

\vspace{-0.4cm}
\subsection{Experiment Design} 
In this section, we describe the design of the experiment to examine these hypotheses. 

\subsubsection{Human-Swarm Simulator} \label{section:Human-Swarm-Simulator}
We extended the Human And Robot Interactive Swarm (HARIS) simulator to run our experiment~\cite{hunt2023demonstrating}. Originally developed for human-swarm interaction~\cite{ramchurn2015study}, HARIS allows the human operator to continuously monitor and control the swarm's behaviour. We selected HARIS over other popular options because it features continuous human interaction with the swarm as a core feature of its engine. HARIS allows us to record and analyze the users' interactions throughout the scenario. We use the search and rescue scenario, modelling a small swarm of drones which must search an area for potential human casualties, discovering potential casualties when they fly near them; this is reported live to the user. The operator's role in this scenario is to designate regions of interest for the swarm to search, and either use the in-built task allocation algorithm (e.g., MaxSum ~\cite{Rogers2011,DelleFave2012}) to suggest an allocation of drones to search regions, or override this and manually assign each drone; these allocations can always be adapted by the operator in real-time (see Fig. \ref{fig:OperatorView}).


 \subsubsection{Human Operator's Role} To explore the dynamics of human-swarm interaction, we used the HARIS simulator to devise an instructive scenario that involved participants managing and controlling a swarm consisting of numerous drones. Our experiment focused on casualty identification, which did not require extensive prior knowledge or training. During the scenario, the swarm provided participants with a series of images, and their objective was to determine whether each image depicted a human casualty or not. To emphasize the significance and potential consequences of accurate classification, we informed participants that these tasks carried a high-risk factor, as an erroneous classification could lead to harm in real-life scenarios. The images used in the experiment were pre-designed and retrieved from a database, without any image processing performed by the robots. They were carefully selected to pose challenges in identifying human casualties amidst other objects in the field, such as trees or animals. This deliberate selection aimed to simulate real-world scenarios where differentiating human casualties from other elements could be difficult. To assess the efficacy of the images in fulfilling their intended purpose, we conducted a manipulation check study involving nine participants. Their task was to determine whether the target in each image was a human or non-human casualty. Each participant completed a total of 24 classification tasks. Our analysis of the results revealed that participants achieved an average accuracy of 64.28\% in completing the classification task. These findings provided valuable insights into the suitability and difficulty level of the images for casualty identification within the context of our scenario.

\begin{figure}[!htb]
  \centering
  \framebox{\includegraphics[width=0.45\textwidth]{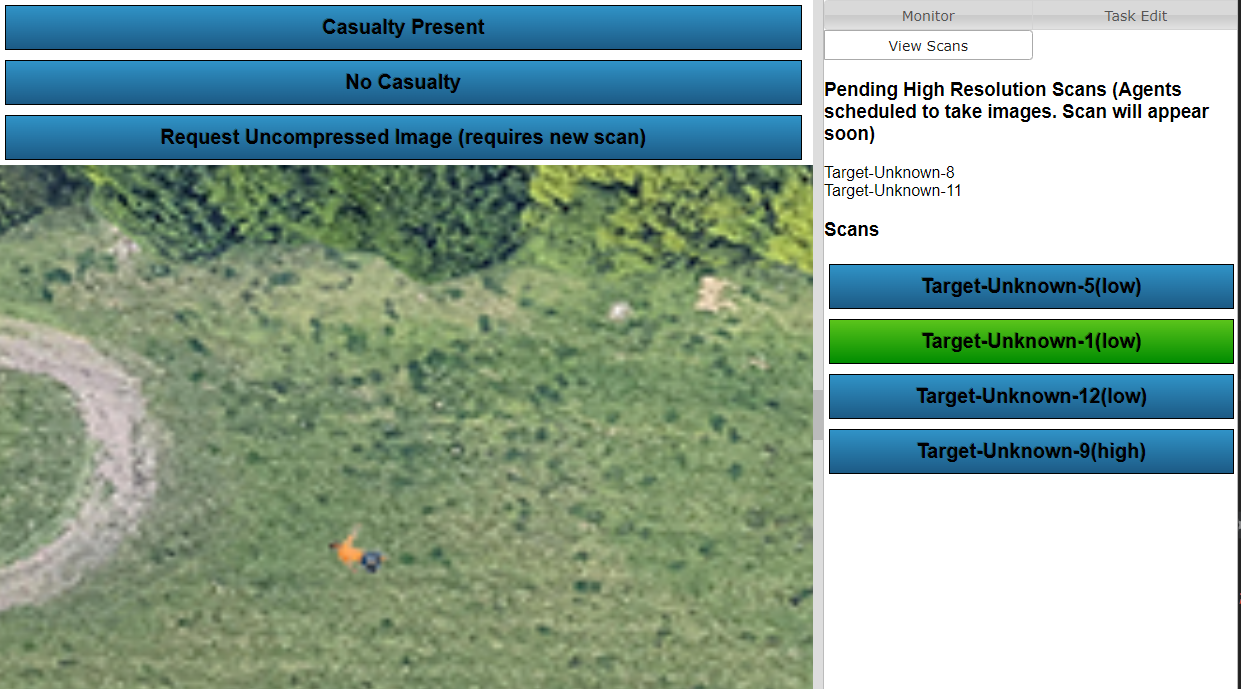}}
  \caption{The scan view shows the target image. The user can classify the target if they are certain of what they see, or can request a better image}
  \label{fig:ScanView}
\end{figure}

\vspace{-0.3cm}
Participants were assigned the role of a swarm operator, allocating search regions for a swarm of Unmanned Aerial Vehicles (UAVs) to find human casualties on a map. The drones automatically return low-resolution images of potential casualties; the participant is required to confirm whether these are real casualties via the buttons on the view tab of the interface as shown in Fig. \ref{fig:ScanView}. Participants in the high data quality group have an extra feature in their interface, a button to request high-resolution images to aid the casualty classification task. Every image is a top-down view of an open area that includes the image of a human or an object edited in. Requesting a high-resolution image results in an unallocated drone being tasked from the drone hub (home position) to the position where the low-resolution image was taken to fetch a higher-resolution scan and return back to the home position before presenting the image scan back to the human for classification. 


 \subsubsection{Study conditions} To examine our hypotheses, the experiment followed a 2x2 mixed design, encompassing both between-subject and within-subject factors. The independent variables included data quality and task workload (density). Participants were randomly assigned to one of two main conditions, namely the high-resolution group or the low-resolution group. Each participant was given an hour to complete the experiment. The experiment consisted of two scenarios: a low workload scenario and a high workload scenario, both of which participants had to complete. In the low task density condition, a set of 8 images was utilized, comprising 5 valid casualties and 3 decoys. On the other hand, the high task density condition involved 12 images, with 7 of them being valid casualties and the remaining 5 serving as decoys. Participants had 8 minutes to complete each task density scenario. Thus, the key distinction between the two task density groups was the relative proportion of decoys. Each participant was expected to locate a total of 20 targets. These targets were distributed within a park located in Southampton, United Kingdom, at coordinates ($50.929544$, $-1.409595$). 

We employed a mixed design for several reasons. Firstly, by using a between-subject design, we aimed to assess the impact of data quality while assuming that any other differences between groups were evenly distributed through random assignment. Thus, we designated data quality as the between-subject factor. Secondly, the within-subject design was chosen to mitigate potential random noise in the data caused by individual characteristics. This was particularly relevant when collecting subjective measures of workload, as a between-subject design could introduce noise. Therefore, we selected workload as the within-subject factor to ensure more precise and reliable results. To minimize potential learning effects and biases, we implemented a counterbalanced study design. This approach helped ensure that half of the participants completed the low workload scenario first followed by the high workload scenario (AB), while the remaining participants underwent the high workload scenario first and then the low workload scenario (BA). By counterbalancing the order of the scenarios, we aimed to reduce the influence of any potential order-based biases and optimize the validity of the findings.

\subsubsection{Measures}
At the end of each condition, we asked participants to complete three scales: a) We used the NASA TLX questionnaire ($6$ questions)~\cite{hart1986nasa} to measure participants' workload after each condition, b) We also used Jian's questionnaire ($7$ questions)~\cite{jian2000foundations} to check participants' trust perception after completing each condition, c) Users also completed a performance questionnaire ($8$ questions). We included two manipulation-check questions to test participants' attention when answering the scales. The behaviour of the users was captured in log files of the HARIS simulator to measure other behavioural aspects such as Human-Swarm performance, interaction with the high-resolution feature, and scenario completion time. 



\subsection{Experiment Procedure} \label{Section:Experiment-Procedure}

The study started with the participants being briefed on what the study was about and a 10 minutes video tutorial hosted on YouTube~\footnote{User study tutorial available online: \url{https://www.youtube.com/watch?v=HhD3zU6jTSQ}}. A three-question validation check was performed to ensure that participants paid attention while watching the tutorial video. This was an essential inclusion criterion for our participants to make sure they understood the scenario. The participants were then provided with the participant information sheet and required to consent in order to continue their participation. After this, the participant demographic details of gender, education level, computer expertise, and familiarity with drone/Swarm Robotics were collected.

Participants were then required to complete a 5-minute tutorial on the online simulator to familiarise themselves with how the simulator works and the type of tasks they would perform. After this, the first scenario of the experiment is loaded. Each scenario interface is similar to that shown in Fig. \ref{fig:OperatorView}. Participants had 8 minutes to complete the scenario. Immediately after completing this, the participants were required to complete the study questionnaire for the scenario as presented in the previous section. They then proceeded to complete the second condition and completed the questionnaire for the second condition. Once this is completed, they were guided to the submission page where they could submit their data and were redirected back to the Prolific platform (online study crowd-sourcing platform~\cite{prolific}) to confirm participation and data submission.


\subsection {Participants Demographic}
We recruited a total of $120$ participants through Prolific. The study was conducted using the cloud-hosted version of the HARIS simulator~\footnote{Online HARIS simulator: \url{https://uos-hutsim.cloud/}}. We received ethics approval from the University of Southampton ethics committee (ERGO number: 69418). All participants were compensated for their time participating in the study. 

The gender distribution of the participant was $75$ male, $41$ female, $3$ non-binary, and $1$ ``preferred not to say''. $74$ of the participants had at least a Bachelor's or a Master's degree. $67$ of the participants were above-average computer users. In terms of UAV/Swarm experience or knowledge, $7$ of the participants had experience in operating drones, $55$ only knew a little about the technology, and the remaining $58$ had no knowledge about drones or Swarm Robotics.

\section{Results}

In this section, we present the quantitative results and annotate each subsection regarding the RQs that they primarily answer. We conducted quantitative analysis on various dependent variables related to participants' scenario performance, behavioural patterns interacting with the swarm, and subjective measures. For each dependent variable, we performed a separate mixed-effects regression with data quality and task density as the fixed effects and participants as the random effects. The regression model further included participants' self-reported swarm knowledge, age group, gender, and qualifications as control variables. We report the descriptive statistics (mean values and standard division) for each quantitative measure and significant results from the regression analysis. On average, participants found and classified $11.5$ targets in $16$ minutes (8 minutes per scenario).

\subsection{Task performance} We started by analysing the measurements reflecting how well participants performed the tasks as described in the Method (Section \ref{Section:Method}), specifically the total number of casualties detected, and the Precision and Recall of their identified problems. Descriptive statistics, including means and standard deviations are presented in Table~\ref{table:performance}.

\begin{table}[h]
\centering
\caption{The mean and standard deviation (in parenthesis) of task performance measures: the number of casualties detected by participants (N casualties) normalised by the number of ground-truth casualties in each condition, the precision and recall of the identified casualties. (Legend: DQ - Data Quality. TD - Task Density.)}
\label{table:performance}
    \begin{tabular}{|c|c|c|c|c|}
        \hline
        \rowcolor{Gray}
          & & N casualties & Precision & Recall \\ 
        \hline
        \multirow{2}{*}{Low DQ} & Low TD & 0.74 (0.08) & 0.750 (0.032) & 0.662(0.024) \\
        \cline{2-5}
        & High TD & 0.64 (0.05) & 0.802 (0.047) & 0.603(0.037) \\
        \hline
        \multirow{2}{*}{High DQ} & Low TD & 0.69 (0.08) & 0.921 (0.056) & 0.632(0.041) \\
        \cline{2-5}
        & High TD & 0.55 (0.04) & 0.872 (0.061) & 0.581(0.039) \\
        \hline
    \end{tabular}
\end{table}

We performed a mixed-effects regression model on the number of identified casualties, Precision, and Recall respectively. For the number of identified casualties (normalised), we found significant two-way interaction between data quality and workload ($B = -0.34$, $SE = 0.15$, $F(2,114) = 5.42$, $p < 0.05$). The posthoc analysis found the contrast between low and high workload marginally significant for high data quality conditions ($p = 0.1$), but not significant for low data quality ($p=0.35$), suggesting that the interactive effect was mainly caused by the delay that requesting more data quality pictures for casualties will result in a lower number of casualties detected. 

While we did not find any significant effect on recall, we found the same significant one-way interaction for data quality on Precision ($B = 0.29$, $SE = 0.06$, $F(2,114) = 5.98$, $p<0.05$). This suggests that data quality indeed increased the quality of the decision (Precision) but does not necessarily increase the number of human casualties (Recall). These two-way interactions on the total number of casualties (human and non-human) and precision implies that when interacting with high-quality images participants found significantly fewer casualties (human and non-human), in low and high workload scenarios.  

\subsection{Behavioural Patterns} 

We have also examined behavioural metrics that reflected how participants interacted with the simulator, including the time spent completing each scenario and the number of times they accessed the high-quality images feature.

\begin{figure}[!htb]
  \centering
    \adjustbox{trim=1cm 0cm 1.2cm 0cm}{
      \includegraphics[width=0.55\textwidth]{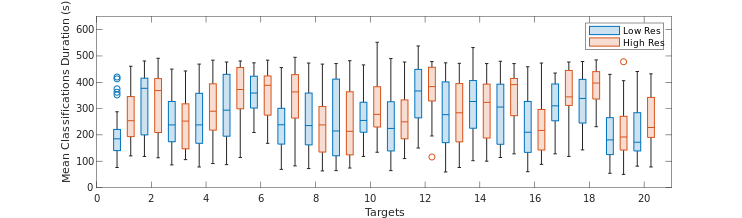}
  }
  \caption{Comparing DQ group target classification duration. Box plot pairs (Low Resolution and High Resolution) show the mean classification duration for each of the 20 targets.}
  \label{fig:target_classification}
\end{figure}

 The distribution of the time spent classifying each of the 20 targets by each study group is shown in Fig. \ref{fig:target_classification}. By aggregating this performance, the mean classification duration for the low data quality group was $265.3$ seconds. The mean classification duration for the high-resolution group was $283.7$ seconds. This shows that the mean classification duration for the low-resolution group was shorter than that of the higher-resolution group by about $18.4$ seconds. A one-way ANOVA was performed to determine if this difference was significant and we see that there was a statistically significant difference in the mean classification duration between the two groups ($F(1, 1374) = 8.675$, $p = 0.003$). The value of F is $8.675$, which reaches significance with a p-value of $0.003$ (which is less than the $0.05$ alpha level).

\begin{table}[h]
\scriptsize
\centering
\caption{Participant groups classification summary with the number of high-resolution image requests in parenthesis. (Legend: DQ - Data Quality. TD - Task Density.)}
\label{table:quantitative_data_summary}
    \begin{tabular}{|c|c|c|c|c|c|}
        \hline
        \rowcolor{Gray}
         Quality & Density & Group Cx & Correct & Incorrect & Fatal \\
        \hline
        \multirow{3}{*}{Low DQ} & Low TD & 284 (-) & 206 (-) & 78 (-) & 42 (-) \\
        \cline{2-6}
        & High TD & 438 (-) & 318 (-) & 118 (-) & 41 (-)\\
        \cline{2-6}
        & Subtotal & 720 (-) & 524 (-) & 196 (-) & 83 (-) \\
        \hline
        \multirow{3}{*}{High DQ} & Low TD & 253 (144) & 207 (119) & 46 (25) & 29 (20)\\
        \cline{2-6}
        & High TD & 403 (167) & 363 (155) & 40 (12) & 20 (11) \\
        \cline{2-6}
        & Subtotal & 656 (311) & 570 (274) & 86 (37) & 49 (31)\\
        \hline
        \rowcolor{Gray}
        \multicolumn{2}{|c|}{Total} & 1376 (311) & 1094 (274) & 282 (37) & 132 (31) \\
        \hline
    \end{tabular}
\end{table}

Table~\ref{table:quantitative_data_summary} provides a quantitative summary of the performance result obtained from the simulator. From this table, a total of $1376$ target classifications were made by the $120$ participants, $720$ by the low data quality group, and $656$ by the higher quality group. The classification accuracy of the low data quality group was $73\%$ and that of the high data quality group was $87\%$. To achieve this level of performance accuracy, the high data quality group requested high-resolution images $47\%$ of the time. Fatal misclassifications are classifications where a valid human casualty has been misclassified as ``not a casualty''. This is a costly misclassification as the consequence is a definite loss of life rather than a mere waste of resources due to the alternative misclassification. The column labelled \textit{Fatal} in Table \ref{table:quantitative_data_summary} records this. The low data quality group made more fatal misclassifications than the high data quality group ($83:49$).

\begin{figure}[!htb]
  \centering
    \adjustbox{trim=1cm 0cm 1.2cm 0cm}{
      \includegraphics[width=0.55\textwidth]{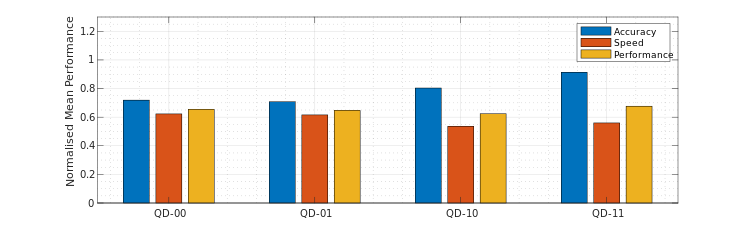}
  }
  \caption{Quantitative performance of the participants grouped by data quality (Q) and task density (D). QD-00: low data quality, low task density. QD-01: low data quality, high task density. QD-10: high data quality, low task density. QD-11: high data quality, high task density.}
  \label{fig:quantitative_performance}
\end{figure}

Breaking down this analysis further, Fig. \ref{fig:quantitative_performance} shows the quantitative performance of the participants. The performance trade-off between speed and accuracy is obvious. The group with lower quality data (QD-00 and QD-01) has lower accuracy but higher speed. The group with the higher quality data (QD-10 and QD-11) had a higher classification accuracy but lower classification speed. The performance bar is the weighted sum of the normalised accuracy ($A_s$) and speed ($R_s$) data. That is, the performance per scenario $ P_s = (A_s * 0.33) + (R_s * 0.67)$. It shows that although the group that had high data quality option perceived their performance to be higher, their performance was actually similar to the low data quality group. Accuracy $A_{s}$ and speed (rate of task completion, $R_{s}$) are computed per scenario as,

\begin{equation}
    A_{s} = \sum_{i=1}^{K} \frac{n_{s,i}}{N_{s,i}} \quad \textrm{and} \quad     R_{s} = \sum_{i=1}^{K} \frac{N_{s,i}}{T_{s,i}}
\end{equation}

Where $K$ is the number of participants in the scenario, $n_{s,i}$ is the number of correct classifications of participant $i$ in scenario $s$, $N_{s,i}$ is the total number of targets classified, $T_{s,i}$ is the scenario completion time which is $8$ minutes by default. The speed $R_s$ for the high task density group QD-01 and QD-11 is normalised by dividing by $1.5$, the standard rate of completing all $12$ classifications in $8$ minutes. The speed $R_s$ for the low task density group QD-00 and QD-10 is normalised by dividing by $1$, the standard rate of completing all $8$ classifications in $8$ minutes.


\begin{figure*}[!htb]
    \centering
    \begin{subfigure}[b]{0.32\textwidth}
        \centering
        \adjustbox{trim=1cm 0cm 1.2cm 0cm}{
          \includegraphics[width=0.99\textwidth]{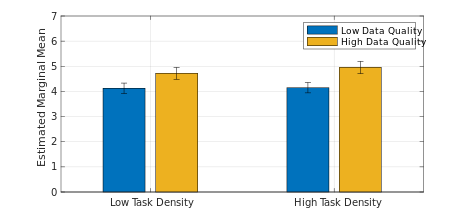}
        }
        \caption{Cognitive workload}
        \label{fig:nasa_tlx_workload}
    \end{subfigure}
    \hfill
    \begin{subfigure}[b]{0.32\textwidth}
        \centering
        \adjustbox{trim=1cm 0cm 1.2cm 0cm}{
            \includegraphics[width=0.99\textwidth]{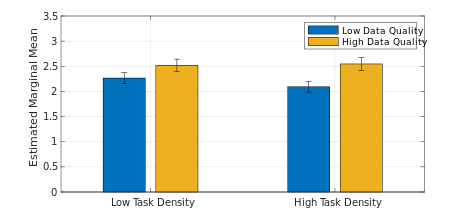}
        }
        \caption{Trust
        }
        \label{fig:jian_trust}
    \end{subfigure}
    \hfill
    \begin{subfigure}[b]{0.32\textwidth}
        \centering
        \adjustbox{trim=1cm 0cm 1.2cm 0cm}{
            \includegraphics[width=0.99\textwidth]{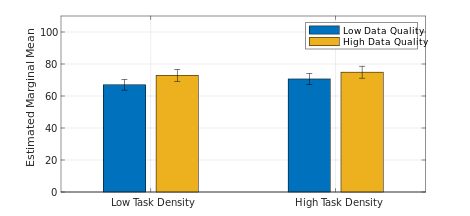}
        }
        \caption{Performance perception
        }
        \label{fig:subjective_performance}
    \end{subfigure}
    \caption{Estimated marginal means of cognitive workload, trust, and performance perception for each of the four conditions studied. a) The cognitive workload is higher for both experiment scenarios performed by the high-resolution group. b) The high-resolution group trusted the system more than the low-resolution group in both of their experiment scenarios. c) The high-resolution group reported a better performance perception than the low-resolution group for both of their experiments.}
    \label{fig:estimated_scenario_marginal_means}
\end{figure*}

\subsection{Subjective perceptions}
In this section, we test the two hypotheses proposed in Section \ref{Section:Method}. We analysed the measurements of the subjective perceptions of the swarm. After completing each casualty detection scenario, participants completed a questionnaire rating their trust perception, workload and performance perception, as described in Section \ref{Section:Experiment-Procedure}.  Table \ref{table:subjective} shows the descriptive statistics of these questionnaire responses. 

\begin{table}[h]
\scriptsize
\centering
\caption{The subjective perception (Trust, workload, and performance) of the swarm in the post-scenario questionnaire, with mean values and standard deviation. (Legend: DQ - Data Quality. TD - Task Density.)}
\label{table:subjective}
    \begin{tabular}{|c|c|c|c|c|}
        \hline
        \rowcolor{Gray}
              & &Trust & Workload & Performance \\
        \hline
        \multirow{3}{*}{Low DQ} & Low TD & 2.266(1.241) & 4.126 (1.805) & 67.03 (22.674) \\
        \cline{2-5}
        & High TD & 2.093(1.204) & 4.150 (1.841) & 70.67(22902) \\
        \cline{2-5}
        & Total DQ &2.180(1.220) & 4.1383 (1.841) & 68.85(22.765) \\
        \hline
        \multirow{2}{*}{High DQ} & Low TD & 2.516(1.279) & 4.720(1.981) & 72.88 (16.014) \\
        \cline{2-5}
        & High TD & 2.5467 (1.368) & 4.953(1.919) & 74.90(17.492) 
        \\
        \cline{2-5}
        &  Total & 2.531(1.319) & 4.8367(1.945) & 73.89(16.729)
        \\
        \hline
    \end{tabular}
\end{table}

Analysis of the Covariance (ANCOVA) test was conducted with Trust, Workload and Performance as the dependent variables; data quality as the independent variable; and gender, age group, and UAV experience as the covariates in order to test the first hypothesis, H1.  Our results show that participants' cognitive workload significantly increased when the data quality increased [Estimated Marginal Mean (EMM) low data quality $= 4.1383 (1.841)$, EMM high data $= 4.8367(1.945)$, $F(1,239) = 8.260$, $P<0.01$]. We also found that participants' trust significantly increased when data quality increased [EMM low data quality $= 2.180(1.220)$, EMM high data $= 2.531(1.319)$, $F(1,239) = 4.593$, $P<0.05$]. A similar pattern was also observed in the data for swarm performance perception [EMM low data quality $= 68.85(22.765)$, EMM high data $= 73.89(16.729)$, $F(1,239) = 3.822$, $p < 0.05$]. All the results obtained were statistically significant, which validates the H1 hypothesis that the presence of high-quality data increases trust perception and overall human-swarm performance. These relationships between data quality and trust, cognitive workload, and performance are summarised graphically in Fig. \ref{fig:estimated_scenario_marginal_means}.

In order to test the second hypothesis, H2, we also conducted ANCOVA with Trust, workload and performance as dependent variables and task density as independent variables. All subjective measures were found non-significant for task density. Therefore, we reject the second hypothesis, H2, that suggested increasing the task density would decrease trust perception and overall human-swarm performance.

\section{Discussion}

The results show that although the performance of both data quality groups was similar, the high-quality data group had a higher sense of achievement, despite the higher cognitive workload reported. This is attributed to the fact that having more features means having more control and being better equipped with the necessary tools needed for precise classification. Our results support this as the group with the higher quality images option used this feature 50\% of the time. This result agrees with the findings in \cite{Yin2019} and \cite{Capiola2022} that the perception of higher accuracy (performance) results in increased trust. Therefore, the availability of high-quality data increases trust and performance perception in human-swarm interaction.

Increasing the task density did not affect trust perception and overall human-swarm performance. The relationship between the subjective measures of trust, workload, and perceived performance against task density was not significant.

The total number of correct classifications between the high and low data quality groups was similar. This implies that procuring sophisticated agents equipped with complex data acquisition sensors may not necessarily lead to higher productivity and the human-swarm interaction experience must always be considered. This observation may have a significant effect on the design of future human-swarm systems. Offering more features and data may lead to less productivity in the swarm operation as the operator may accomplish fewer tasks and spend more time on accuracy. Swarm operators with simpler autonomous agents would accomplish more tasks, cover wider areas, and complete the mission more efficiently. 

Higher priority to accuracy may be justified due to the cost of resource allocation where the number of available ground rescue teams is limited. In this case, accuracy becomes a critical factor and there is a need to be certain of the casualty before dispatching a rescue team. Therefore, to assure certainty, the cost of procuring expensive autonomous agents equipped with more features may be justifiable. Fatal misclassifications are false negatives where the operator identifies a casualty as non-casualty and non-fatal classifications are false positives where non-casualties are marked as human casualties. Both cases can have significant consequences. In fatal cases, the issue is clear but also in non-fatal cases, it can be argued that they are as dangerous due to suboptimal resource utilization. If the available rescue team is limited, the cost of investigating a false casualty target means there is less time to reach a true casualty and save a life. The balance between requesting a high volume of data with low accuracy or a low volume of data with high accuracy could be dynamically modelled and implemented in human-swarm search and rescue missions. This means that when resources are abundant, false negatives can be investigated but as resources become depleted, a slower but more accurate approach is desired to increase resource utilization.

\section{Conclusion}

In this paper, we investigated the effect of data visualisation quality (presence of high quality images) and task density (workload) in human-swarm interaction. We conducted a user study using the online HARIS simulator with 120 participants. The operators allocated to search a grid and classified images  captured by the UAVs in the simulation. We find that data quality has almost no effect on the number of successful classifications but the high data quality leads to a higher accuracy rate, as expected. The group with access to low-quality data was able to classify more targets which reinforces the common speed vs accuracy trade-off issue. The availability of high-quality data increases trust and performance perception in human-swarm interaction. It was shown that although the performance of both data quality groups was similar, the high-quality data group had a higher sense of achievement despite the higher cognitive workload reported, regardless of the actual scenario performance. 


The trust and cognitive workload relationship to data visualisation quality and task density presented in this work relies on subjective data provided by experiment participants. There is a need to corroborate this with objective tools that measure trust and cognitive workload more quantitatively. Also, the ideal nature of our simulation environment does not take into account the real-world conditions (such as battery consumption, weather, etc.) that may affect the UAVs' performance in their flight or when capturing photos. Future works would consider using brain signals to estimate cognitive workload more accurately. In addition to this, field studies that replicate real-world conditions would be required to decrease the reality gap. 

\section*{APPENDIX}
\subsection*{List of Abbreviations}
\vspace{-0.3cm}
\begin{table}[!h]
\label{table:list_of_abbr}
    \begin{tabular}{ll}
            ANCOVA & Analysis of the Covariance \\
            DQ & Data Quality \\
            EMM & Estimated Marginal Mean \\
            HARIS & Human And Robot Interactive Swarm \\
            TD & Task Density \\
            UAV & Unmanned Aerial Vehicle \\
    \end{tabular}
\end{table}

\vspace{-0.1cm}
\section*{ACKNOWLEDGMENT}

The authors wish to thank Dr Miguel Massot-Campos and Ashley Pare for their contributions during the conceptualisation of this project. This project was done as part of the EPSRC Smart Solutions Towards Cellular-Connected Unmanned Aerial Vehicles System (EP/W004364/1) and Explainable Human-swarm Systems project funded by UKRI Trustworthy Autonomous Systems Hub (EP/V00784X/1).  

\bibliographystyle{IEEEtran}
\bibliography{IEEEabrv,ref}

\end{document}